**Global Smart Spaces**

# Working Document on Gloss Ontology (Revised)

## D9 [PART 2]

*J. COUTAZ, A. DEARLE, S. DUPUY-CHESSA, G.N.C. KIRBY, C. LACHENAL, R. MORRISON, G. REY, E. ZIRINTSIS*

| | | | |
|---|---|---|---|
| **IST Project Number** | IST-2000-26070 | **Acronym** | GLOSS |
| **Full title** | Global Smart Spaces | | |
| **EU Project officer** | Thomas Skordas | | |

| | | | | | |
|---|---|---|---|---|---|
| **Deliverable** | **Number** | D9 | **Name** | | |
| **Task** | **Number** | T | **Name** | | |
| **Work Package** | **Number** | WP | **Name** | | |
| **Date of delivery** | **Contractual** | | **Actual** | | |
| **Code name** | | | **Version** 1.0 draft ☐ final ☑ | | |
| **Nature** | Prototype ☐ Report ☑ Specification ☐ Tool ☐ Other: | | | | |
| **Distribution Type** | Public ☐ Restricted ☑ to: <partners> | | | | |
| **Authors (Partner)** | J. Coutaz, A. Dearle, S. Dupuy-Chessa, G.N.C. Kirby, C. Lachenal, R. Morrison, G. Rey, E. Zirintsis | | | | |
| **Contact Person** | G.N.C. Kirby | | | | |
| | **Email** | graham@dcs.st-and.ac.uk | **Phone** | | **Fax** |
| **Abstract (for dissemination)** | This document describes the Gloss Ontology. The ontology and associated class model are organised into several packages. Section 2 describes each package in detail, while Section 3 contains a summary of the whole ontology. | | | | |
| **Keywords** | | | | | |

# 1 INTRODUCTION

This document describes the Gloss Ontology. The ontology and associated class model are organised into several packages. Section 2 describes each package in detail, while Section 3 contains a summary of the whole ontology.

# 2 PACKAGES

## 2.1 UNIVERSE

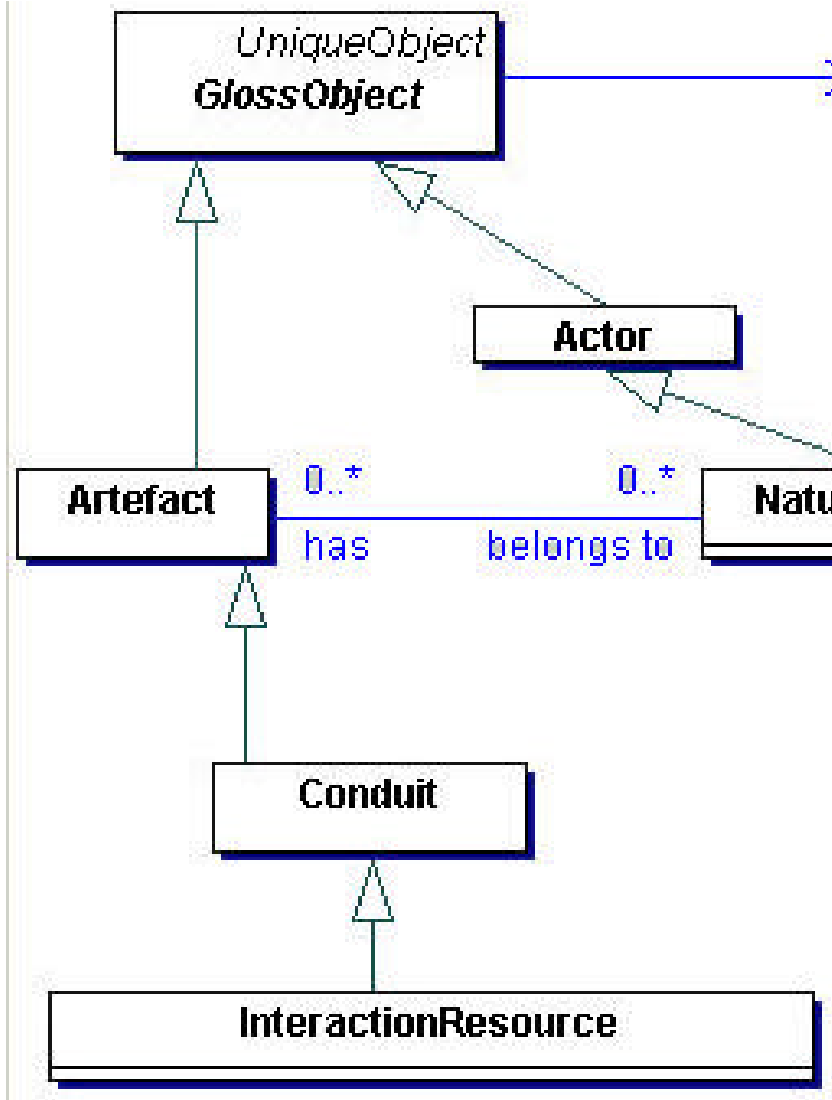

| Class Summary | |
|---|---|
| Activity | Activity that someone or something is doing. |
| Actor | Representation of a Gloss-enabled actor. |
| Aeroplane | A mode of transport. |
| Artefact | An inanimate entity that is significant in the GLOSS universe. |
| Bicycle | A mode of transport. |
| Car | A mode of transport. |
| Conduit | A distinguished artefact that acts as a conduit for information transfer with the GLOSS fabric. |
| Foot | A mode of transport. |
| GlossObject | Superclass unifying the concepts of Actor and Artefact. |
| Information | Arbitrary data in arbitrary format. |
| ModeTransport | A mode of transport. |
| Profile | A profile associated with a GLOSS-enabled person. |
| Train | A mode of transport. |

## 2.2 INTERACTION

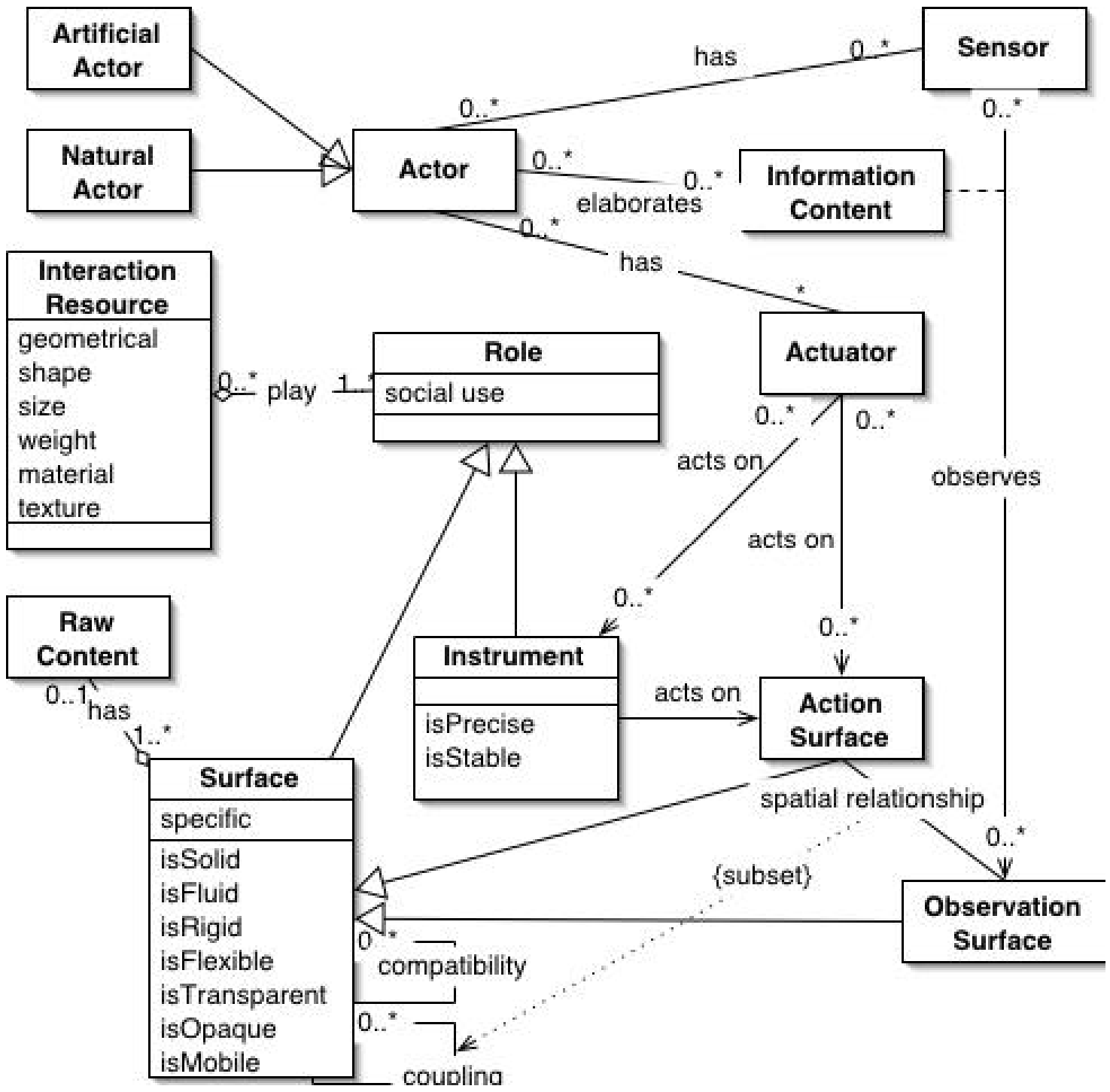

| Class Summary | |
|---|---|
| ActionSurface | A subset of a physical surface on which an actor can act directly with actuators and/or indirectly with instruments. |
| Actor | A system or person. |
| Actuator | A component that is used by actors to modify the state of an interaction resource. |
| ArtificialActor | A system. |
| InformationContent | Information drawn upon by actors to perform computation, built from many sources including raw contents that can be observed from a surface. |
| Instrument | An object held in the hand by a natural actor to operate a user interface. |
| InteractionResource | A mediator between an artificial actor and a natural actor, which may serve as an Instrument and/or as a Surface. |
| NaturalActor | A person. |
| ObservationSurface | A subset of a physical surface that an actor can observe with sensors. |
| **RawContent** | Observed raw contents of a surface. |
| **Role** | Superclass unifying the concepts of Surface and Instrument. |
| **Sensor** | A component that is used by actors to observe the state of interaction resources. |
| **Surface** | A physical surface. |

### 2.2.1 ATTRIBUTES

| Attributes | |
|---|---|
| **Surfaces** | Because multi-surface interaction is grounded on physicality, we suggest the following (non-exhaustive) list of attributes to characterize surfaces: the geometrical shape (e.g., sphere, polygon, a human face as in Hypermask where computer-generated expressions are projected on a blank mask), size and weight, material (e.g., wood, plastic, vapor, water), color, texture (e.g., homogeneous and smooth), and social use (public, private). Attributes of a surface determine the modalities that are necessary for observing and for acting: Modalities for observation denote the sensors involved in observing the content of a surface. For human actors, these include sight, hearing, etc. Modalities for action denote the classes of actions that are applicable to the surface such as writing, folding, and moving. |
| **Instruments** | Attributes of an instrument are grounded on physicality: shape, size, weight, material, social use (private, sharable, etc.). From these attributes, one infers its modalities for observation (e.g., touch, sight) as well as its modalities for action (e.g., physical/digital ink scribbling, pointing, moving, reshaping, illuminating). |

### 2.2.2 PROPERTIES

| Properties | |
|---|---|
| **Surfaces** | A property is the capability of an entity to fulfill a particular function. In HCI, properties provide a useful structure for design.<br>- The *solidity/fluidity/nebulosity* of a surface is inferred from the material it is made of. Liquid or vapor, it is ephemeral but it can be traversed. For example, a rain curtain on which images are projected, serves as a passage between virtual and real worlds.<br>- The *rigidity/flexibility* of a surface expresses its capacity to change its shape and size. Most interactive surfaces are rigid. However, the electronic paper, from Xerox and MIT, open the way to foldable surfaces. Illuminating Clay, which allows the user to shape a surface made of clay, is another promising approach to non-rigid interactive surfaces.<br>- The *opacity/transparency* of a surface is widely exploited in civil architecture. A transparent surface enriches itself with environmental information. It favours openness while forming a boundary. Windowpanes augmented with piezo-electric transducers allow pedestrians to interact with the shop by tapping.<br>- *Mobility* coupled with *lightness* and *smallness* opens the way to new usage. A lightweight small size surface like a PDA can be carried in the hand. If so, it may also serve as a pointing instrument.<br>- *Writability/erasibility* denotes the capacity of a surface to be modified with “write actions” and its capacity to be erased. A public wall is writable but tagging a wall is socially incorrect: it cannot be erased in a simple way. On the other hand, light and sound shows, public-animated walls based on tracking human movements are |
| | appropriate since they use digital ink, an erasable material.<br>- *Heterogeneity* may enforce partitioning of a surface into areas for actions and areas for observation. For example, the whiteboard area of the GLOSS table affords scribbling, the wooden-look plastic area suggests piling up private paper documents, and the central bright circular area serves as the primary focus for interaction with the system. However, these hypotheses about the affordance of surface heterogeneity need to be verified.<br>- *Refraction* and *reflexion* of a surface have a direct impact on its *observability*. In particular, shiny surfaces are nightmare for computer vision based sensors. *Reachability* denotes whether the surface is physically accessible directly: too high, a surface may not be used for action or may require a dedicated instrument (e.g., a ladder or a laser pointer). |
| **Instruments** | Properties of an instrument are measured against the functions expected from the instrument. For example, *precision* and *stability* are relevant properties for pointing instruments. Scribbling is concerned with *manipulability*. |

### 2.2.3 *SPATIAL RELATIONSHIPS*

| **Spatial Relationships** | |
|---|---|
| **Surfaces, Actors** | The purpose of a topology is to describe the location and orientation of entities in a reference coordinate system. Here, the entities of interest are those of our ontology, i.e., the actors, actuators, sensors, instruments, observation and action surfaces involved in a particular interactive situation.<br>The user's position in a multi-surface space matters. For example, the GLOSS circular area can be rotated on user's request by clicking on the white border of the circle. Alternatively, this articulatory task could migrate to the system if GLOSS were able to maintain the location of the user with regard to the table.<br>Similarly, the relative positions of surfaces matter. The orientation of the rendering surfaces relative to the user (e.g., "horizontal" and "vertical") determines the nature of the output modalities.<br>In the Pick-and-Drop example, the user brings a PDA close to an electronic wall-board. In this configuration, the user can pick any information from the PDA and drop it on the board at the location denoted by the PDA position. This capacity to detect proximity provides one way to control surface coupling.<br><br><u>Coupling</u> between surfaces denote their mutual dependency. Two surfaces are coupled when a change of state of one surface has an impact on the state of the other. The Pick-and-Drop presented is one such example.<br><u>Compatibility</u> between surfaces expresses the possibility to use them conjointly, by complementarity, redundancy, equivalence, and assignation.<br>- The painter metaphor is an example of multi-surface interaction based on *complementarity*: tools palettes are displayed on the PDA, the PDA is held in the non dominant hand, the user holds a stylus in the dominant hand, and like the painter artist picks the appropriate tool on the palette with the stylus, then draws on the canvas supported by the wall-size electronic screen. In this example, the complementarity between the PDA and the electronic board is grounded on the differences between their size and weight attributes. Complementarity between surfaces may also rely on the similitude of their attributes to define new functions.<br>- Surfaces are composed in a *redundant* way when they are used simultaneously to accomplish the same task. For example, connecting a Smart Board to the video output of a PC allows users to duplicate the user interface on both the electronic board and the display screen of the PC.<br>- Surfaces are functionally *equivalent* when they can be used alternatively to accomplish a given set of tasks. For example, it is increasingly popular to access web services through a workstation, a PDA, or a cellular phone.<br>- Surface *assignment* means that each surface of the configuration plays a particular |
| | role. For example, in a meeting, personal information editing is assigned to PDA's whereas collaborative editing of a document is assigned to the public electronic board. |
| **Instruments** | As for surfaces, instruments belong to the topology mentioned earlier, and their relationships can be analysed according to their level of *coupling* and *compatibility*.<br>For example in the UJF Magic table, two tokens are coupled by bringing them into contact. When coupled, the user can select physical/digital markings by forming a rectangle with the two tokens, one in each hand. Selected physical markings are digitised. Then, selected markings can be simultaneously resized and rotated using the tokens. Coupling ends when one of the tokens is hidden with the hand. |

### 2.2.4 COUPLING WITH CONTENT

Spatio-temporal coupling and generality/specificity of coupling are concerned with associating interaction resources to information content.

- **Spatio-temporal Coupling with Content:** spatio-temporal coupling between entities defines how these entities are associated in time and space. To illustrate the discussion, let us consider the coupling of instruments and actuators to content (whether it is raw content or informational). Fitzmaurice's observed that in conventional GUIs, the coupling of instruments to logical functions (i.e., information content) is "time-multiplexed": there is only one such instrument attached at a time. As a result, instruments are repeatedly coupled and decoupled to content. With Graspable UIs, the coupling can be "space-multiplexed": different instruments can be attached to different content, each independently (but possibly simultaneously) accessible to the user.

  As indicated by our ontology, actuators, such as fingers, can be coupled directly to content without any intermediate instrument. As a generalization of Fitzmaurice's work, we suggest to consider the numbers of actuators, sensors, instruments, and surfaces that can be simultaneously coupled to information content. These numbers can be used as a metric to characterize the system capabilities in relation to human performance and needs. For example, the Magic Table, which is able to track more than 4 tokens at a time, makes possible multi-user interaction.

- **Genericity/Specifity of Coupling with Content:** an interaction resource is generic if it can be coupled to any type of content. The mouse, Fitzmaurice's bricks, The Magic Table tokens, and the GLOSS clicker are generic instruments. The GLOSS table and wall are generic surfaces. Alternatively, an interaction resource is specific when it is dedicated to a particular class of content. Phicons, and more specifically the MIT Dome developed for the Metadesk are specific instruments. In HyperMask, masks which have the shape of a human face, are specific surfaces.

  Specificity works by analogy with real world entities. As demonstrated by early work on physical programming (and many others), specificity facilitates understanding and exploration. On the other hand, they are not reusable. They cannot satisfy the scale factor when the number of information types increases. Therefore, the right mix of generic/specific interaction resources needs to be identified in relation to information space and finality of the system.

## 2.3 SPACE

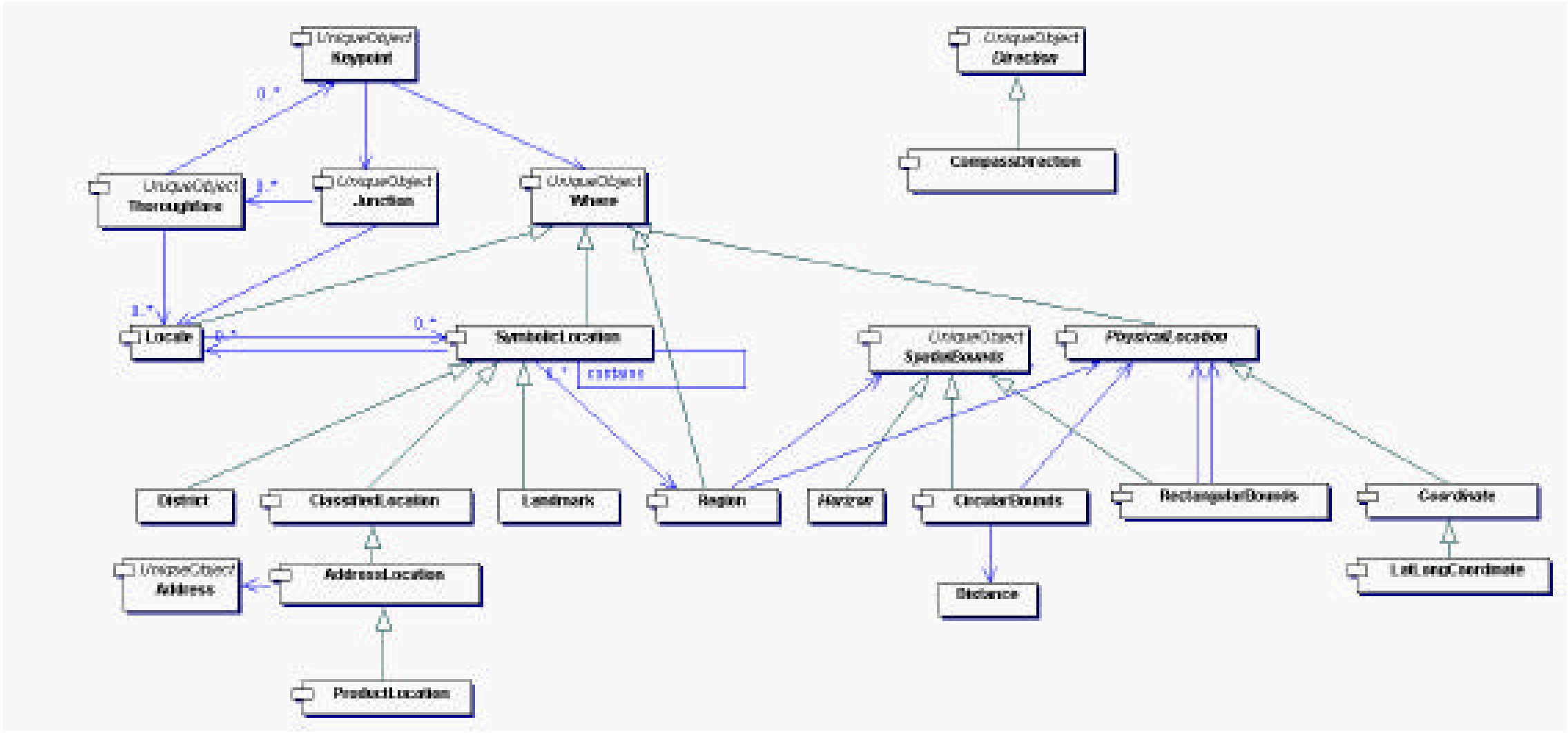

| **Class Summary** | |
|---|---|
| **Address** | A postal address. |
| **AddressLocation** | A ClassifiedLocation with an Address. |
| **CircularBounds** | A 2-D circular bounding region. |
| **ClassifiedLocation** | A SymbolicLocation annotated with additional information. |
| **CompassDirection** | A Direction represented as a bearing in degrees from True North. |
| **Coordinate** | A point represented in some geographical coordinate system, which may be 2-dimensional or 3-dimensional. |
| **Direction** | An orientation represented in some way. |
| **Distance** | Representation of the distance between two Wheres. |
| **District** | A district within a town, city or country. |
| **Horizon** | Region currently perceived by a Gloss user. |
| **Junction** | A junction at which a number of Thoroughfares meet. |
| **Keypoint** | An individual point on a Thoroughfare. |
| **Landmark** | A well-known landmark. |
| **LatLongCoordinate** | A 2-dimensional point represented as degrees latitude and longitude. |
| **Locale** | A logical grouping of SymbolicLocations. |
| **PhysicalLocation** | A point represented in some way. |
| **ProductLocation** | An AddressLocation where a particular user service may be obtained. |
| **RectangularBounds** | A 2-D rectangular bounding region. |
| **Region** | A bounded fixed region of space that may contain or intersect with other regions. |
| **SpatialBounds** | A 2-dimensional or 3-dimensional shape bounding a region. |
| **SymbolicLocation** | An entity, fixed or moveable, that may contain people, artefacts and other locations. |
| **Thoroughfare** | A sequence of Keypoints and related Locales. |
| **Where** | Superclass unifying the concepts of PhysicalLocation, Region and SymbolicLocation. |

### 2.3.1 XML REPRESENTATION

The spatial ontology can also be represented as an XML document. Generating XML is required when a location event message has to be transmitted from one artificial actor to another. The XML document conforms to the schema shown below (schema can also be found at http://www-systems.dcs.st-and.ac.uk/gloss/xml/2003-07/locationEvent.xsd):

```
<?xml version="1.0" encoding="iso-8859-1" ?>
<xsd:schema elementFormDefault="qualified"
  xmlns=http://www-systems.dcs.st-and.ac.uk/gloss/xml/2003-07/
  targetNamespace="http://www-systems.dcs.st-and.ac.uk/gloss/xml/2003-07/"
  xmlns:xsd="http://www.w3.org/2001/XMLSchema">
   <xsd:include schemaLocation="http://www-systems.dcs.st-
                                and.ac.uk/gloss/xml/2003-07/spaceModel.xsd" />
  <xsd:element name="locationEvent" type="LocationEvent" />
  <xsd:complexType name="LocationEvent">
    <xsd:sequence>
     <!-- ******* Mandatory elements. ******* -->
     <!-- ID of user or artefact being observed; -->
     <xsd:element name="ID" type="ID" />
     <!-- List of previous processing steps for this event; -->
     <xsd:element name="processingSequence">
       <xsd:complexType>
         <xsd:sequence>
           <xsd:element    name="processingStep"
                           maxOccurs="unbounded" minOccurs="0">
            <xsd:complexType>
              <xsd:sequence>
                <xsd:element name="dateTime" type="xsd:dateTime" />
                <xsd:element name="description" type="xsd:string" />
              </xsd:sequence>
            </xsd:complexType>
           </xsd:element>
         </xsd:sequence>
        </xsd:complexType>
     </xsd:element>
     <!-- One or more location observations. -->
     <xsd:element name="observation" minOccurs="1" maxOccurs="unbounded">
       <xsd:complexType>
           <xsd:sequence>
             <!-- *** General location information.**** -->
             <!-- The time at which the observation was made. -->
             <xsd:element name="timeOfObservation" type="xsd:dateTime" />
             <!-- The location. -->
             <xsd:element name="where" type="Where" />
             <!-- ****** Optional elements from here onward. ***** -->
```

```
            <!-- Altitude. -->
            <xsd:element minOccurs="0" maxOccurs="1" name="altitude"
                                                 type="Altitude" />
            <!-- Speed. -->
            <xsd:element minOccurs="0" maxOccurs="1" name="speed"
                                                 type="Speed" />
            <!-- Course: direction of current travel (true bearing). -->
            <xsd:element minOccurs="0" maxOccurs="1" name="course"
                                                 type="Bearing" />
            <!-- Current magnetic variation from true north. -->
            <xsd:element minOccurs="0" maxOccurs="1" name="magneticVariation"
                                                 type="Bearing" />
            <!-- ****** GPS-specific information. ****** -->
            <!-- The number of satellites visible, from 0 to 12 inclusive. -->
            <xsd:element minOccurs="0" maxOccurs="1" name="satellitesVisible">
               <xsd:simpleType>
                 <xsd:restriction base="xsd:integer">
                    <xsd:minInclusive value="0" />
                    <xsd:maxInclusive value="12" />
                 </xsd:restriction>
               </xsd:simpleType>
            </xsd:element>
            <!-- Dilution of precision. -->
            <xsd:element minOccurs="0" maxOccurs="1" name="PDOP"
                                                 type="xsd:float" />
            <!-- Horizontal dilution of precision. -->
            <xsd:element minOccurs="0" maxOccurs="1" name="HDOP"
                                                 type="xsd:float" />
            <!-- Vertical dilution of precision. -->
            <xsd:element minOccurs="0" maxOccurs="1" name="VDOP"
                                                 type="xsd:float" />
            <!-- Estimated horizontal error in metres. -->
            <xsd:element minOccurs="0" maxOccurs="1" name="HPE"
                                                 type="xsd:float" />
            <!-- Estimated vertical error in metres. -->
            <xsd:element minOccurs="0" maxOccurs="1" name="VPE"
                                                 type="xsd:float" />
          </xsd:sequence>
        </xsd:complexType>
      </xsd:element>
     </xsd:sequence>
   </xsd:complexType>
 </xsd:schema>
```

The various different XML representations of the spatial ontology which are also used in the above schema are defined as shown below (the type definitions can be found at http://www-systems.dcs.st-and.ac.uk/gloss/xml/2003-07/spaceModel.xsd):

```
<?xml version="1.0" encoding="ISO-8859-1" ?>
<xsd:schema elementFormDefault="qualified"
  xmlns="http://www-systems.dcs.st-and.ac.uk/gloss/xml/2003-07/"
  targetNamespace="http://www-systems.dcs.st-and.ac.uk/gloss/xml/2003-07/"
  xmlns:xsd="http://www.w3.org/2001/XMLSchema">
  <xsd:complexType name="Where">
     <xsd:choice minOccurs="0" maxOccurs="1">
       <xsd:element name="symbolicLocation" type="SymbolicLocation" />
       <xsd:element name="physicalLocation" type="PhysicalLocation" />
       <xsd:element name="region" type="Region" />
       <xsd:element name="locale" type="Locale" />
     </xsd:choice>
     <xsd:attribute name="name" type="xsd:string" use="optional" />
     <xsd:attribute name="glossURN" type="xsd:anyURI" use="optional" />
```

```
  </xsd:complexType>
  <xsd:complexType name="SymbolicLocation">
     <xsd:sequence>
       <xsd:choice minOccurs="0" maxOccurs="1">
          <xsd:element name="classifiedLocation" type="ClassifiedLocation" />
          <xsd:element name="landmark" type="xsd:string" />
          <xsd:element name="district" type="xsd:string" />
       </xsd:choice>
       <xsd:element name="information" type="Information" />
       <xsd:element name="region" type="Region" />
       <xsd:element name="locale" minOccurs="0" maxOccurs="unbounded"
                                                type="Locale" />
       <xsd:element name="fixed" type="xsd:boolean" />
     </xsd:sequence>
  </xsd:complexType>
  <xsd:complexType name="PhysicalLocation">
     <xsd:choice minOccurs="0" maxOccurs="1">
       <xsd:element name="coordinate" type="Coordinate" />
     </xsd:choice>
  </xsd:complexType>
  <xsd:complexType name="Region">
     <xsd:sequence>
       <xsd:element name="distinguishedPoint" type="PhysicalLocation" />
       <xsd:element name="bounds" type="SpatialBounds" />
     </xsd:sequence>
  </xsd:complexType>
  <xsd:complexType name="SpatialBounds">
     <xsd:choice minOccurs="0" maxOccurs="1">
       <xsd:element name="horizon" type="xsd:string" />
       <xsd:element name="circularBounds" type="CircularBounds" />
       <xsd:element name="rectangularBounds" type="RectangularBounds" />
     </xsd:choice>
  </xsd:complexType>
  <xsd:complexType name="CircularBounds">
     <xsd:sequence>
       <xsd:element name="centre" type="PhysicalLocation" />
       <xsd:element name="radius" type="Distance" />
     </xsd:sequence>
  </xsd:complexType>
  <xsd:complexType name="RectangularBounds">
     <xsd:sequence>
       <xsd:element name="topLeft" type="PhysicalLocation" />
       <xsd:element name="bottomRight" type="PhysicalLocation" />
     </xsd:sequence>
  </xsd:complexType>
  <xsd:complexType name="Coordinate">
     <xsd:choice minOccurs="0" maxOccurs="1">
          <xsd:element name="latLongCoordinate" type="LatLongCoordinate" />
     </xsd:choice>
  </xsd:complexType>
  <xsd:complexType name="LatLongCoordinate">
     <xsd:sequence>
          <xsd:element name="latitude" type="Latitude" />
          <xsd:element name="longitude" type="Longitude" />
     </xsd:sequence>
  </xsd:complexType>
  <!-- Latitude in degrees from -90 to 90 inclusive. Negative latitudes are in
the southern hemisphere; positive latitudes are in the northern hemisphere.-->
  <xsd:simpleType name="Latitude">
     <xsd:restriction base="xsd:double">
          <xsd:minInclusive value="-90" />
          <xsd:maxInclusive value="90" />
     </xsd:restriction>
  </xsd:simpleType>
```

```
  <!-- Longitude in degrees from -180 to 180 inclusive. Negative longitudes
are west of Greenwich; positive longitudes are east of Greenwich. -->
  <xsd:simpleType name="Longitude">
     <xsd:restriction base="xsd:double">
          <xsd:minInclusive value="-180" />
          <xsd:maxInclusive value="180" />
     </xsd:restriction>
  </xsd:simpleType>
  <!-- Altitude in specified units (M:metres or F:feet, default metres). -->
  <xsd:complexType name="Altitude">
     <xsd:simpleContent>
          <xsd:extension base="xsd:double">
               <xsd:attribute default="M" name="unit" use="optional">
                    <xsd:simpleType>
                         <xsd:restriction base="xsd:string">
                              <xsd:enumeration value="F" />
                              <xsd:enumeration value="M" />
                         </xsd:restriction>
                    </xsd:simpleType>
               </xsd:attribute>
          </xsd:extension>
     </xsd:simpleContent>
  </xsd:complexType>
  <xsd:complexType name="Distance">
          <xsd:simpleContent>
               <xsd:extension base="NonNegativeDouble">
                    <xsd:attribute default="m" name="unit" use="optional">
                         <xsd:simpleType>
                              <xsd:restriction base="xsd:string">
                                   <xsd:enumeration value="m" />
                                   <xsd:enumeration value="km" />
                                   <xsd:enumeration value="miles" />
                                   <xsd:enumeration value="nautical miles" />
                              </xsd:restriction>
                         </xsd:simpleType>
                    </xsd:attribute>
               </xsd:extension>
          </xsd:simpleContent>
  </xsd:complexType>
  <xsd:complexType name="Speed">
     <xsd:simpleContent>
          <xsd:extension base="NonNegativeDouble">
               <xsd:attribute default="knots" name="unit" use="optional">
                    <xsd:simpleType>
                         <xsd:restriction base="xsd:string">
                              <xsd:enumeration value="m/s" />
                              <xsd:enumeration value="km/h" />
                              <xsd:enumeration value="miles/h" />
                              <xsd:enumeration value="knots" />
                         </xsd:restriction>
                    </xsd:simpleType>
               </xsd:attribute>
          </xsd:extension>
     </xsd:simpleContent>
  </xsd:complexType>
  <!-- Bearing from 0 to 360 inclusive. -->
  <xsd:simpleType name="Bearing">
     <xsd:restriction base="xsd:double">
          <xsd:minInclusive value="0" />
          <xsd:maxInclusive value="360" />
     </xsd:restriction>
  </xsd:simpleType> <!-- Used for quantities that cannot be negative. -->
  <xsd:simpleType name="NonNegativeDouble">
     <xsd:restriction base="xsd:double">
          <xsd:minInclusive value="0" />
```

```
      </xsd:restriction>
   </xsd:simpleType>
   <!-- Email address must contain at least one character before an '@', and at
least one '.' within the following domain. -->
   <xsd:simpleType name="EmailAddress">
      <xsd:restriction base="xsd:string">
           <xsd:pattern value="[^@]+@[^.]+\..+" />
      </xsd:restriction>
   </xsd:simpleType>
   <!-- Phone number must begin with '+' followed by only digits or spaces. -->
   <xsd:simpleType name="PhoneNumber">
      <xsd:restriction base="xsd:string">
           <xsd:pattern value="\+[0-9 ]*" />
      </xsd:restriction>
   </xsd:simpleType>
   <!-- ID of user or artefact being observed: arbitrary bitstring which may or
may not be unique. -->
   <xsd:complexType name="ID">
      <xsd:choice>
           <xsd:element name="bitString" type="xsd:string" />
           <!-- Globally unique ID of user or artefact being observed. -->
           <xsd:element name="GUID" type="xsd:string" />
           <!-- Phone number of user or artefact being observed. -->
           <xsd:element name="phone" type="PhoneNumber" />
           <!-- Email address of user or artefact being observed. -->
           <xsd:element name="email" type="EmailAddress" />
      </xsd:choice>
   </xsd:complexType>
   <xsd:complexType name="Information">
      <xsd:sequence>
           <xsd:element maxOccurs="unbounded" minOccurs="0" name="info"
                                                    type="xsd:string" />
           <xsd:element maxOccurs="unbounded" minOccurs="0" name="link"
                                                    type="xsd:anyURI" />
      </xsd:sequence>
   </xsd:complexType>
   <xsd:complexType name="Classification">
      <xsd:sequence>
           <xsd:element maxOccurs="unbounded" minOccurs="1"
                              name="classificationType" type="xsd:string" />
      </xsd:sequence>
   </xsd:complexType>
   <xsd:complexType name="Address">
      <xsd:sequence>
           <xsd:element minOccurs="0" maxOccurs="1" name="nameNumber"
                                               type="xsd:string" />
           <xsd:element minOccurs="0" maxOccurs="1" name="street"
                                               type="xsd:string" />
           <xsd:element minOccurs="0" maxOccurs="1" name="town"
                                               type="xsd:string" />
           <xsd:element minOccurs="0" maxOccurs="1" name="county"
                                               type="xsd:string" />
           <xsd:element minOccurs="0" maxOccurs="1" name="postCode"
                                               type="xsd:string" />
           <xsd:element minOccurs="0" maxOccurs="1" name="webAddress"
                                               type="xsd:anyURI" />
           <xsd:element minOccurs="0" maxOccurs="1" name="email"
                                               type="EmailAddress" />
      </xsd:sequence>
   </xsd:complexType>
   <xsd:complexType name="Locale">
      <xsd:sequence>
           <xsd:element maxOccurs="1" minOccurs="0" name="parent"
                                               type="Locale" />
           <xsd:element maxOccurs="unbounded" minOccurs="0"
```

```
                          name="classification" type="Classification" />
            <xsd:element maxOccurs="unbounded" minOccurs="0" name="contents"
                                                    type="SymbolicLocation" />
            <xsd:element maxOccurs="unbounded" minOccurs="0" name="neighbours"
                                                    type="Locale" />
            <xsd:any maxOccurs="unbounded" minOccurs="0" />
        </xsd:sequence>
    </xsd:complexType>
    <xsd:complexType name="ClassifiedLocation">
        <xsd:sequence>
            <xsd:choice minOccurs="0" maxOccurs="1">
                <xsd:element name="addressLocation" type="AddressLocation" />
            </xsd:choice>
            <xsd:element maxOccurs="unbounded" minOccurs="0"
                          name="classification" type="Classification" />
            <xsd:element name="description" type="xsd:string" />
        </xsd:sequence>
    </xsd:complexType>
    <xsd:complexType name="AddressLocation">
        <xsd:sequence>
            <xsd:choice minOccurs="0" maxOccurs="1">
                <xsd:element name="productLocation" type="ProductLocation" />
            </xsd:choice>
            <xsd:element name="address" type="Address" />
        </xsd:sequence>
    </xsd:complexType>
    <xsd:complexType name="ProductLocation">
        <xsd:sequence>
            <xsd:element name="openTime" type="xsd:time" />
            <xsd:element name="closeTime" type="xsd:time" />
        </xsd:sequence>
    </xsd:complexType>
</xsd:schema>
```

Below are examples of XML documents that conform to the schema specified above and represent various location events encapsulating the spatial ontology. Each of the messages makes different use of the spatial ontology: the first two use a coordinate, whereas the third uses a region.

***Example 1***

```
<?xml version="1.0" encoding="ISO-8859-1"?>
<locationEvent xmlns=http://www-systems.dcs.st-and.ac.uk/gloss/xml/2003-07/
  xmlns:xsi=http://www.w3.org/2001/XMLSchema-instance
  xsi:schemaLocation="http://www-systems.dcs.st-and.ac.uk/gloss/xml/2003-07/
      http://www-systems.dcs.st-and.ac.uk/gloss/xml/2003-07/locationEvent.xsd">
      <ID>
        <email>graham@dcs.st-and.ac.uk</email>
      </ID>
      <processingSequence></processingSequence>
      <observation>
            <timeOfObservation>2003-05-16T18:31:59</timeOfObservation>
            <where>
                  <physicalLocation>
                        <coordinate>
                              <latLongCoordinate>
                                    <latitude>56.340232849121094</latitude>
                                    <longitude>-2.86754378657099878</longitude>
                              </latLongCoordinate>
                        </coordinate>
                  </physicalLocation>
            </where>
      </observation>
```

```
</locationEvent>
```

***Example 2***

```
<?xml version="1.0" encoding="ISO-8859-1"?>
<locationEvent xmlns="http://www-systems.dcs.st-and.ac.uk/gloss/xml/2003-07/"
  xmlns:xsi="http://www.w3.org/2001/XMLSchema-instance"
  xsi:schemaLocation="http://www-systems.dcs.st-and.ac.uk/gloss/xml/2003-07/
     http://www-systems.dcs.st-and.ac.uk/gloss/xml/2003-07/locationEvent.xsd">
     <ID>
       <phone>+447941615809</phone>
     </ID>
     <processingSequence>
          <processingStep>
               <dateTime>2003-05-16T18:31:59</dateTime>
               <description>processed on PDA</description>
          </processingStep>
     </processingSequence>
     <observation>
          <timeOfObservation>2003-05-16T18:31:59</timeOfObservation>
          <where>
               <physicalLocation>
                    <coordinate>
                         <latLongCoordinate>
                              <latitude>56.340232849121094</latitude>
                              <longitude>-2.867543786570998789</longitude>
                         </latLongCoordinate>
                    </coordinate>
               </physicalLocation>
          </where>
          <altitude unit ="F">123.45</altitude>
          <speed>35.1</speed>
          <course>45</course>
          <magneticVariation>3.8</magneticVariation>
          <satellitesVisible>05</satellitesVisible>
          <PDOP>1.5</PDOP>
          <HDOP>1.5</HDOP>
          <VDOP>1.5</VDOP>
          <HPE>15.0</HPE>
          <VPE>15.0</VPE>
     </observation>
</locationEvent>
```

***Example 3***

```
<?xml version="1.0" encoding="ISO-8859-1"?>
<locationEvent xmlns="http://www-systems.dcs.st-and.ac.uk/gloss/xml/2003-07/"
  xmlns:xsi="http://www.w3.org/2001/XMLSchema-instance"
  xsi:schemaLocation="http://www-systems.dcs.st-and.ac.uk/gloss/xml/2003-07/
     http://www-systems.dcs.st-and.ac.uk/gloss/xml/2003-07/locationEvent.xsd">
     <ID>
       <bitString>graham</bitString>
     </ID>
     <processingSequence>
          <processingStep>
               <dateTime>2003-05-16T18:31:59</dateTime>
               <description>processed on PDA</description>
          </processingStep>
          <processingStep>
               <dateTime>2003-05-16T18:32:01</dateTime>
               <description>routed through node 18B6</description>
          </processingStep>
```

```
          <processingStep>
               <dateTime>2003-05-16T18:32:02.42</dateTime>
               <description>received on server</description>
          </processingStep>
     </processingSequence>
     <observation>
          <timeOfObservation>2003-05-16T18:31:59</timeOfObservation>
          <where>
               <region>
                    <distinguishedPoint>
                         <coordinate>
                              <latLongCoordinate>
                                   <latitude>56.340232849121094</latitude>
                                   <longitude>-2.8675437865709987</longitude>
                              </latLongCoordinate>
                         </coordinate>
                    </distinguishedPoint>
                    <bounds>
                     <circularBounds>
                      <centre>
                       <coordinate>
                        <latLongCoordinate>
                         <latitude>56.3402328491210234</latitude>
                         <longitude>-2.8675437865709945453</longitude>
                        </latLongCoordinate>
                       </coordinate>
                      </centre>
                      <radius unit="miles">1</radius>
                     </circularBounds>
                    </bounds>
               </region>
          </where>
          <altitude>123.45</altitude>
          <satellitesVisible>05</satellitesVisible>
          <PDOP>1.5</PDOP>
     </observation>
</locationEvent>
```

## 2.4 TIME

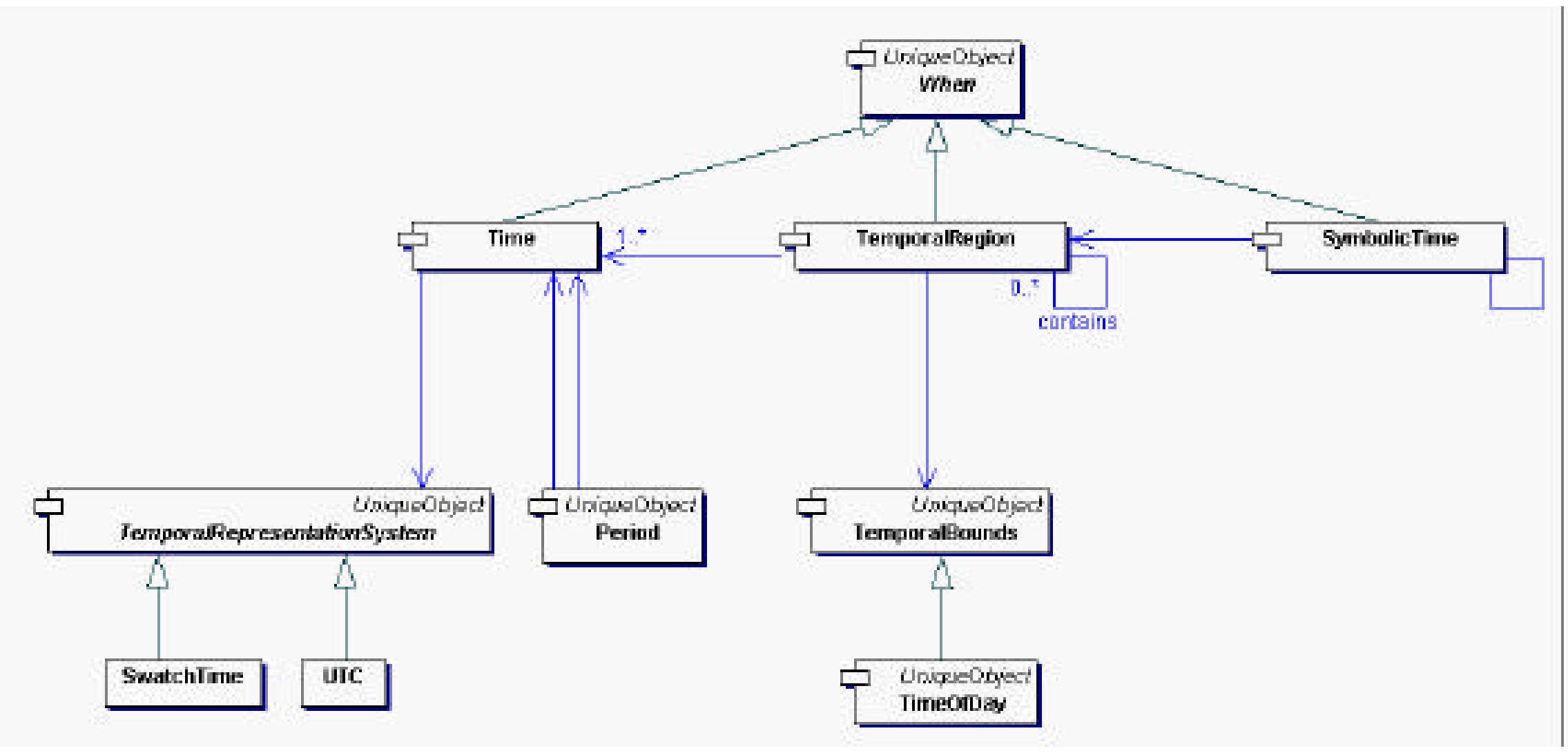

| Class Summary | |
|---|---|
| **Period** | A contiguous period of time. |
| **SwatchTime** | Swatch temporal system. |
| **SymbolicTime** | A time instant, period or set of periods. |
| **TemporalBounds** | Temporal analogy to SpatialBounds: a set of one or more Periods. |
| **TemporalRegion** | A set of time periods. |
| **TemporalRepresentationSystem** | A system for representing time. |
| **Time** | A point in time expressed using a particular representation system. |
| **TimeOfDay** | A time of day expressed without reference to any particular day. |
| **UTC** | UTC temporal system. |
| **When** | Superclass unifying the concepts of Time, SymbolicTime, and TemporalRegion. |

## 2.5 METAPHORS

### *2.5.1 CLASS DIAGRAM*

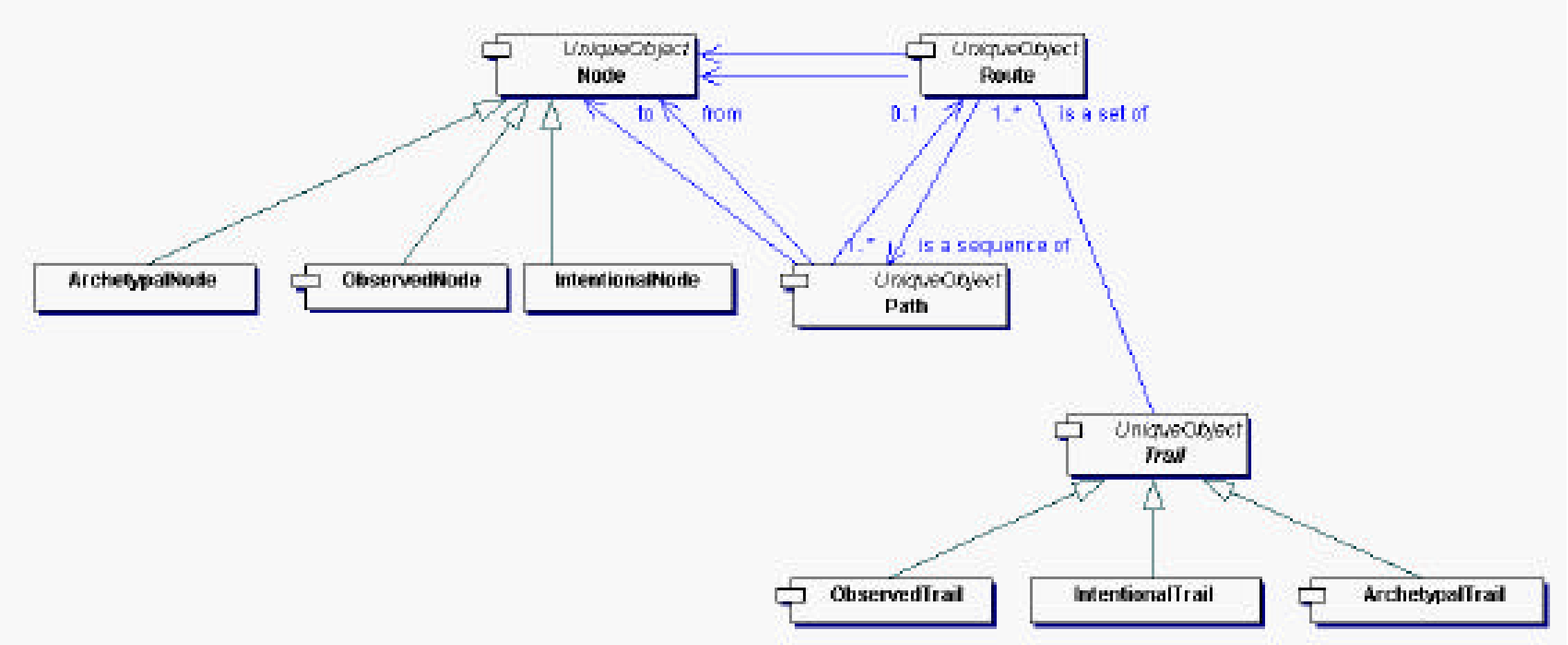

| Class Summary | |
|---|---|
| **ArchetypalNode** | An individual node within an archetypal trail. |
| **ArchetypalTrail** | A route between two Wheres, with optional multiple sub-routes. |
| **IntentionalNode** | An individual node within an IntentionalTrail. |
| **IntentionalTrail** | An unordered set of locations, regions or coordinates, with associated information, linked by theme. |
| **Node** | An individual node within an ArchetypalTrail or IntentionalTrail. |
| **ObservedNode** | An individual observation within an ObservedTrail. |
| **ObservedTrail** | An ordered sequence (a snail trail) of observations of a Person or Artefact, each recording a time (a When), a place (a Where) and optionally some additional information. |
| **Path** | A path between two Wheres. |
| **Route** | A sequence of Paths linking a start Where and an end Where. |
| **Trail** | Abstract superclass for various trail classes. |

# 3 COMBINED ONTOLOGY

| | |
|---|---|
| **ActionSurface** | A subset of a physical surface on which an actor can act directly with actuators and/or indirectly with instruments. |
| **Activity** | Activity that someone or something is doing. |
| **Actor** | A system or person. |
| **Actuator** | A component that is used by actors to modify the state of an interaction resource. |
| **Aeroplane** extends ModeTransport | A mode of transport. |
| **Address** | A postal address. |
| **AddressLocation** extends ClassifiedLocation | A *ClassifiedLocation* with an *Address*. |
| **ArchetypalNode** | An individual node within an *ArchetypalTrail*. |
| **ArchetypalTrail** extends Trail | A directed graph of locations, regions or coordinates, together with associated information and a recommended order for visiting them. The trail is a graph rather than a sequence since it may contain alternative sub-routes, for example to cater for different modes of transport or other user preferences. Example instances include:<br>• a route from St Andrews to Grenoble;<br>• a recommended order to visit Scottish Whisky Distilleries.<br><br>An archetypal trail might be distilled from a number of observational trails. It has no intrinsic time dimension, although information about travel times between nodes could be included. |
| **Artefact** extends GlossObject | An inanimate entity that is significant in the GLOSS universe. Example instances include:<br>• a building;<br>• a road junction;<br>• a projection screen. |
| **ArtificalActor** | A system. |
| **Bicycle** extends ModeTransport | A mode of transport. |
| **Bounds** | A 2-dimensional or 3-dimensional shape bounding a region. Example instances include:<br>• a regular shape;<br>• an arbitrary polygon |
| **Car** extends ModeTransport | A mode of transport. |
| **CircularBounds** extends SpatialBounds | A 2-D circular bounding region. |
| **ClassifiedLocation** extends SymbolicLocation | A *SymbolicLocation* annotated with additional information. |
| **CompassDirection** extends Direction | A *Direction* represented as a bearing in degrees from True North. |
| **Conduit** extends Artefact | A distinguished artefact that acts as a conduit for information transfer with the GLOSS fabric. A conduit may optionally be associated with a person. Example instances include:<br><br>• a PDA;<br>• a mobile phone;<br>• a car radio;<br>• a display screen;<br>• a Java button. |
| **Coordinate** extends Where | A point represented in some geographical coordinate system, which may be 2-dimensional or 3-dimensional. Example instances include:<br><br>• the position of the first tee on the Old Course, St Andrews, expressed as [latitude, longitude, height];<br>• grid reference 781490 on British Ordnance Survey sheet 51. |
| **Direction** | An orientation represented in some way. |
| **Distance** | Representation of the distance between two *Where*s. |
| **District** extends SymbolicLocation | A district within a town, city or country. |
| **Foot** extends ModeTransport | A mode of transport. |
| abstract **GlossObject** | Superclass unifying the concepts of *Person* and *Artefact*. It represents any identity that is part of or tracked within the GLOSS fabric. A *GlossObject* may be interrogated for its most recent known position. |
| public **Horizon** | Region currently perceived by a Gloss user. |
| **Information** | Arbitrary data in arbitrary format. |
| **InformationContent** | Information drawn upon by actors to perform computation, built from many sources including raw contents that can be observed from a surface. |
| **Instrument** | An object held in the hand by a natural actor to operate a user interface. |
| **IntentionalNode** | An individual node within an *IntentionalTrail*. |
| **IntentionalTrail** extends Trail | An unordered set of locations, regions or coordinates, with associated information, linked by theme. There may be a number of ordered routes through the set. Example instances include:<br><br>• a Scotch Whisky Trail<br>• a Fife Tourist Trail<br>• the sites of the GLOSS consortium<br><br>An intentional trail has no intrinsic time dimension, although information about travel times between nodes could be included. |
| **InteractionResource** | A mediator between an artificial actor and a natural actor, which may serve as an Instrument and/or as a Surface. |
| **Junction** | A junction at which a number of *Thoroughfare*s meet. |
| **Keypoint** | An individual point on a *Thoroughfare*. |
| **Landmark** extends SymbolicLocation | A well-known landmark. |
| **LatLongCoordinate** extends Coordinate | A 2-dimensional point represented as degrees latitude and longitude. |
| **Locale** | A logical grouping of *SymbolicLocation*s. Example instances include:<br>• a whisky locale in Speyside;<br>• a tourism locale in North East Fife. |
| abstract **ModeTransport** extends Information | A mode of transport. Example instances include:<br>• car;<br>• train;<br>• aeroplane;<br>• bicycle;<br>• on foot. |
| **NaturalActor** | A person. |
| **Node** | An individual node within an *ArchetypalTrail* or *IntentionalTrail*. Example instances include:<br>• location Edinburgh Airport;<br>• coordinate British OS NM278162, which is the position of the entrance to the Talisker Distillery. Whisky has been produced here since 1834. |
| **ObservationSurface** | A subset of a physical surface that an actor can observe with sensors. |
| **ObservedNode** extends Node | An individual observation within an *ObservedTrail*. Example instances include:<br>• Graham Kirby was at location Sabena SN2069 at 14:17 UTC on 2nd May 2001, at which point drinks were served.<br>• Sabena SN2069 was at coordinate 2.34E 51.92N 8000m at 15:17 Central European Time on 2/5/01 |
| **ObservedTrail** extends Trail | An ordered sequence (a snail trail) of observations of a *Person* or *Artefact*, each recording a time, a place (coordinate, region or location) and optionally some additional information. Example instances include:<br>• an observational trail recording the travel of a person from St Andrews to Grenoble;<br>• an observational trail recording the travel of a train along a route on a particular day.<br>The individual observations comprising an observational trail might be recorded automatically at fixed time or spatial intervals; manually by user intervention; and/or automatically whenever the observed entity comes into proximity with designated places. |
| **Path** | Represents a path between two *Where*s. |
| **Period** | A contiguous period of time. |
| **PhysicalLocation** extends Where | A point represented in some way. |
| **ProductLocation** extends AddressLocation | An *AddressLocation* where a particular user service may be obtained. |
| **Profile** | A profile associated with a GLOSS-enabled person. Example attributes may include:<br>• current mode of transport; e.g. on foot, car, hovercraft, etc.;<br>• food preferences;<br>• personal interests. |
| **RawContent** | Observed raw contents of a surface. |
| **RectangularBounds** extends Bounds | A 2-D rectangular bounding region. |
| **Region** extends Where | A bounded fixed region of space that may contain or intersect with other regions. A region may be 2-dimensional or 3-dimensional. It is fixed relative to some geographical coordinate system. Example instances include:<br>• the bounds of the Old Course, St Andrews;<br>• the triangle bounded by Dublin, Glasgow and St Andrews;<br>• the volume occupied by the Livingstone Tower, Strathclyde. |
| **Role** | Superclass unifying the concepts of Surface and Instrument. |
| **Route** | A sequence of *Path*s linking a start *Where* and an end *Where*. |
| **Sensor** | A component that is used by actors to observe the state of interaction resources. |
| **SpatialBounds** | A 2-dimensional or 3-dimensional shape bounding a region. |
| **Surface** | A physical surface. |
| **SwatchTime** | Swatch temporal system. |
| **SymbolicLocation** extends Where | An entity, fixed or moveable, that may contain people, artefacts and other locations. A location occupies a region, which may vary over time for a moveable location. Example instances include:<br>• a train (moveable);<br>• a car (moveable);<br>• a bus-stop (fixed);<br>• a room (fixed);<br>• an airport (fixed);<br>• a logical thing filling a defined 3-D space. |
| **SymbolicTime** extends When | A time instant, period or set of periods. |
| **TemporalBounds** | Temporal analogy to *SpatialBounds*: a set of one or more *Periods*. |
| **TemporalRegion** extends When | A set of time periods. |
| **TemporalRepresentationSystem** | A system for representing time. |
| **Thoroughfare** | A sequence of *Keypoint*s and related *Locale*s. |
| **Time** extends When | A point in time expressed using a particular representation system. |
| **TimeOfDay** | A time of day expressed without reference to any particular day. |
| **Trail** | Abstract superclass for various trail classes. |
| **Train** extends ModeTransport | A mode of transport. |
| **UTC** | UTC temporal system. |
| abstract **When** | Superclass unifying the concepts of *Time*, *SymbolicTime*, and *TemporalRegion*. |
| abstract **Where** | Superclass unifying the concepts of *Coordinate*, *Region* and *Location*. The rationale is that in some contexts we may be interested in any one of:<br>• a point position;<br>• a physical fixed region;<br>• a logical location.<br>For example, depending on the underlying hardware technology a person might be tracked to a point (e.g. GPS), a region (e.g. GSM) or a location (e.g. active badge in a building or train). |